\documentclass[useAMS,usenatbib,usegraphicx]{mn2e}
 
\title[Galaxy Merger Time-Scales]{The Structures of Distant Galaxies - IV: A New Empirical Measurement of the Time-Scale for Galaxy Mergers - Implications for the Merger History}
\author[Christopher J. Conselice]{Christopher J. Conselice$^{1}$\thanks{E-mail:
conselice@nottingham.ac.uk} \\
$^{1}$University of Nottingham, School of Physics \& Astronomy, Nottingham, NG7 2RD UK }

\def\solm{M$_{\odot}\,$}

\def\solm{M$_{\odot}\,$}

\def\casgm20{CAS-G-M$_{20}\,$}
\def\m20{M$_{20}\,$}
\begin{document}

\date{Accepted ; Received ; in original form}
\pagerange{\pageref{firstpage}--\pageref{lastpage}} \pubyear{2002}

\maketitle

\label{firstpage}

\begin{abstract}

Understanding the role of mergers in galaxy formation is 
one of the most outstanding problems in extragalactic astronomy.
While we now have an idea for how the merger fraction evolves
at redshifts $z < 3$,  converting this merger
fraction into merger rates, and therefore how many mergers an average
galaxy undergoes during its history, is still uncertain.  
The main reason for this is that the inferred number of mergers depends 
highly upon the time-scale observational methods are sensitive for finding
ongoing or past mergers.  While there are several 
theoretical and model based estimates of merger times,
there is currently no empirical measure of this time-scale.  
We present the first observationally based measurement of
merger times utilising the observed decline in the galaxy major 
merger fraction
at $z < 1.2$ based on $> 20,000$ galaxies in the Extended Groth
Strip and COSMOS surveys.  Using a new methodology described in this paper, we 
are able to determine how long a galaxy remains identifiable as a
merging system within the CAS system.  We find a maximum CAS major merger time-scale
of 1.1$\pm0.3$ Gyr at $z < 1.2$, and a most likely CAS merger time-scale of 
0.6$\pm0.3$ Gyr, in good agreement with results from N-body simulations.  
Utilizing this time-scale we are able to measure the number of major 
mergers galaxies with masses M$_{*} > 10^{10}$ \solm 
undergo
at $z < 1.2$, with a total number N$_{\rm m} = 0.90_{-0.23}^{+0.44}$.  
We further show that this time-scale is 
inconsistent with a star formation origin for ultra-high asymmetries, 
thereby providing further evidence that structural 
methods are able to locate mostly merging galaxies.

\end{abstract}


\section{Introduction}

One of the key observable quantities for understanding the evolution of
galaxies is the merger rate. This is defined as the number of galaxies
which are merging per unit time, per unit volume, as a function of the
history of universe. The merger rate, when known, reveals the ultimate
importance of mergers within the galaxy formation process.
Previous papers in this series have explored this issue in great
depth, and have derived the merger history in terms of the merger
fraction evolution up to $z = 6$ (Conselice et al. 2008; 2009a,b).  These
papers find that the merger fraction increases with higher redshifts,
and largely flattens at roughly $z > 1.5$ or so.  The merger fraction is also
non-neglectable, and in general between 2-6 major mergers occur per
galaxy with stellar mass M$_{*} > 10^{10}$ \solm since $z = 3$.

This large uncertainty in the cumulative number of mergers is the result 
of uncertainties in the time-scale for the merger process. While it is 
relatively easy to measure the merger fraction, converting this into a merger
rate requires knowledge of the time sensitivity of our methods
for locating merging galaxies. For example, if a method of finding
mergers was sensitive to long-lived features, such as
outer tidal tails, then the merger fraction would be high. However,
because the time-scale is also large in this case, the merger rate 
$\sim f_{\rm m}/\tau_{\rm m}$,
would be the same if the process for measuring the merger fraction had
a very short time sensitivity, as while merger fractions can differ,
the merger rate must be the same if measuring the same process (e.g.,
Conselice et al. 2009b).

The major merger fractions and rates we measure in Papers I-III all use the 
CAS system, which is well calibrated on nearby and high-redshift objects in
various environments (e.g., Conselice 
et al. 2000a,b; Conselice 2003; Conselice et al. 
2003; Conselice et al. 2005; Conselice 2006a; 
Conselice et al. 2008; Conselice et al. 2009a,b).  However, a major
unknown factor is the time-scale sensitivity of the CAS system within the
major merger process.  

There have been several attempts to understand and calculate what
this merger time-scale sensitivity is.  The first attempt was
carried out by Conselice (2006a) who measured N-body models of the merger
process to determine how long a merging galaxy would be asymmetric enough
to be considered a major merger.  Conselice (2006a) calculated an average
value of $\tau_{\rm m} = 0.4\pm$0.2 Gyr.  However, these N-body models are 
basic, and do not
include star formation, dust, etc. which are all important
factors in understanding
and interpreting galaxy structure.  More recently, Lotz et al. (2008) 
examined elaborate
galaxy mergers that include dust and star formation, and determined
a merging time-scale for the CAS method of between 0.4 and 1.2 Gyr, depending
on the orbital and initial conditions.  While we have a 
rough range of CAS time-scales, this factor
of two to three uncertainty does not allow us to accurately measure the total
number of mergers, or the merger rate, as a function of time.  We also
have the issue that these time-scales are all model dependent, and it 
remains possible that the actual merger time-scales are different, and
possibly much longer.

In this paper we develop and utilise a new empirical method for measuring  
merger time-scales based on the evolution of the merger fraction at low
redshifts.  We have recently found that the merger fraction declines
at redshifts $z < 1$, and we  use this to obtain a minimum time-scale for 
detecting mergers.  In conclusion, we find that the average CAS merger
time scale at $z < 0.7$ is 0.6$\pm 0.3$ Gyr for galaxies with stellar
masses M$_{*} > 10^{10}$ \solm. We furthermore use this 
value to argue
that the minimum number of major mergers a galaxy with stellar
mass M$_{*} > 10^{10}$ \solm undergoes at $z < 1.2$ is N$_{\rm m}
= 0.90^{+0.44}_{-0.23}$.  We also describe how our results 
constrain the orbital properties of mergers, as well as providing further
evidence that our methods for locating mergers are not strongly biased by
star formation. 
We use a standard cosmology of H$_{0} = 70$ km s$^{-1}$ Mpc$^{-1}$, and 
$\Omega_{\rm m} = 1 - \Omega_{\lambda}$ = 0.3 throughout.

\section{Method}

The basic method for measuring the merger time-scale consists of assuming
that between two relatively close redshift bins, there are no 
additional mergers occurring (we examine and revise this assumption later 
in this paper).  Based on this, the decline in the observed merger
fraction ($f_{\rm m}$), which is simply the number of galaxies which match our
CAS profile for mergers ($N_{\rm m}$, see Papers I-III) divided by the 
total number of
galaxies within the sample being examined ($N_{\rm T}$), is as a function
of redshift and stellar mass, given by: 

\begin{equation}
f_{\rm m}({\rm M_{*}},z) = \frac{N_{\rm m}}{N_{T}}.
\end{equation}

\noindent As an example, at a redshift $z_1$ we observe a CAS merger fraction 
of $f_{\rm 1,m}$.  At a redshift $z_2 < z_1$ we observe a CAS merger fraction 
$f_{\rm 2,m}$.  If no new mergers occur between $z_1$ and $z_2$, 
then the merger
fraction $f_{2,m} \leq f_{1,m}$.  This is due to the fact that no new
mergers are occurring, and we only witness passively evolving morphological
structures. An observed
decline in the merger fraction then results from galaxies which were asymmetric
enough to be within the CAS merger criteria, cooling morphologically to
regular symmetrical systems.   Note that for this calculation we are only 
considering mergers
which are detectable with CAS.  From N-body models and empirical
observations, it appears that gas-rich major mergers are likely the majority
of the types detected (e.g., Conselice 2006; Lotz et al. 2008).  
We furthermore make the assumption that all gas rich major mergers are
detectable with CAS (see Conselice et al. 2006, 2009b).

If we assume that within the population of asymmetric galaxies,
the typical merger time-scale is $\tau_{\rm m}$, and that these asymmetric 
galaxies began their merger sensitivity evenly within the past $\tau_{\rm m}$ 
years previous to $z = z_{1}$, this implies that
at $z_1$ there is a uniform probability that a given asymmetric galaxy 
started its merger process some-time in the past $\tau_{\rm m}$ Gyr.
Therefore after an amount of time $\delta t$  a fraction
of these systems will no longer be asymmetric enough to be counted
as a merger. This number is given by:

\begin{equation}
  N_{\rm m}  \times \frac{\delta t}{\tau_{\rm m}},
\end{equation}

\noindent where, as an extreme example, after the entire merger sensitivity
time-scale has occurred (i.e., $\delta t = \tau_{\rm m}$), there would
remain no galaxies satisfying the merger criteria.
After a time $\delta t$, which we take to be at the observed
redshift range $z = z_2$, the number of observable ongoing mergers is
then, 

\begin{equation}
N_{\rm 2,m} = N_{\rm m} - N_{\rm m} \times \frac{\delta t}{\tau_{\rm m}}.
\end{equation}

\noindent Again, this assumes that there have been no additional mergers
within the mass range of interest between $z_1$ and $z_2$.  If we divide
this equation by $N_{\rm tot}$ we get

\begin{equation}
f_{\rm 2,m} = f_{\rm 1,m} - f_{\rm 1,m}  \times \frac{\delta t}{\tau_{\rm m}},
\end{equation}
 
\noindent where we have assumed that the total number of galaxies has not
significantly changed between redshifts. This equation can then be solved 
to find the value of $\tau_{\rm m}$,

\begin{equation}
\tau_{\rm m} = \delta t \times (1 - \frac{f_{\rm 2,m}}{f_{\rm 1,m}})^{-1}
\end{equation}

\noindent We note that this measurement of $\tau_{\rm m}$ is a `typical' CAS
merger time-scale sensitivity - it is the sum over all merging systems
in their various orbital configurations, dust content, star formation rates,
inclination, etc. For a mixture of systems with different time-scales,
eq. (5) will tend to overestimate the {\rm average} CAS merger time-scale at 
a maximum of 20\%-30\% based on Monte Carlo simulations.  Eq. (5) also does 
not hold when there are mergers 
occurring between the two redshifts of interest. Thus, we cannot use 
equation (5) to measure merger time-scales when merging is ongoing.  

The time-scale from Eq. (5) does however apply to any distribution of
galaxies, even if during a merger a galaxy is asymmetric 
at two or more distinct times.  We test this by a Monte Carlo
simulation of galaxies being asymmetric for two different times, and
derive similar time-scales as for continuous mergers, using eq. (5).  
The derived CAS
merger time-scale does increase slightly if the gap between when a
galaxy remains asymmetric is $>$50\% of the total merger time-scale,
but this is a very unlikely scenario (Lotz et al. 2008).  

We also carry out a Monte Carlo simulation
to determine how much influence a small (10\%) fraction of galaxies
in our sample with long CAS time-scales ($> 10$ Gyr) would affect our results, 
finding no significant increase in the measured time-scale. 
However, simulations suggest that
the range of time-scales differs by only a factor of 2-3, and a wide
time-scale distribution is an unlikely scenario.

We however know that massive galaxies at $z < 1.2$ still
undergo some merging (e.g.,
Conselice et al. 2008; Bluck et al. 2009), even between two
close redshifts.  If we
denote the fraction of galaxies which have begun to undergo
a merger between $z_1$ and $z_2$ as $f_{\rm 2,new}$ then the
time-scale $\tau_{\rm m}$ can be measured as,

\begin{equation}
\tau_{\rm m} = \delta t \times (1 - \frac{(f_{\rm 2,m} - f_{\rm 2,new})}{f_{\rm 1,m}})^{-1}.
\end{equation}

\noindent where the value of $f_{\rm 2,new}$ needs to be determined
independently of the change of merger fraction between $z_1$ and
$z_2$.  We explore possible values for $f_{\rm 2,new}$, although the
value is unlikely to be higher than a few \% at $z < 0.7$.  This approach
can be generalised by the fact that the value of $f_{\rm 2,new}$ depends
on the merger rate, or the inverse merger rate per galaxy, $\Gamma$,
which is measured as $\Gamma = \tau_{\rm m}/f_{\rm gm}$, and the 
evolution of $\Gamma$ can be parameterised as a function of redshift (Conselice
et al. 2008; Bluck et al. 2008).  

If we measure $\Gamma$ in units of 1 Gyr, such that $\Gamma = \Gamma_{\rm 1 gyr} \left(\frac{\tau_{\rm m}}{\rm 1\, Gyr}\right)$, then by considering the 
fraction of new mergers, which is $f_{\rm 2,new} = \delta t/\Gamma$, then
the time-scale $\tau_{\rm m}$ is:

\begin{equation}
\tau_{\rm m} = \delta t \times \frac{(f_{1,{\rm m}} - \Gamma^{-1}_{1\, {\rm gyr}})}{(f_{1,{\rm m}} - f_{2,{\rm m}})},
\end{equation}

\noindent where the value of $\Gamma_{1\, {\rm gyr}}$ is measured using a 
time-scale of 1 Gyr.   The values of $\Gamma$
in this paper are measured,
in a slightly different way than previously, 
as the time between successive mergers.  The value
of $\Gamma$ we use in equation (7) is based on the merger fractions
from Conselice et al. (2009b) (\S 3).





\section{Data}

The data we use for this paper originates from the study of 
merger fractions at $ z < 1.2$, published by Conselice et al. (2009b).
Conselice et al. (2009b) 
find that the merger fraction drops steadily at $z < 1.2$, although the
most rapid drop is seen at $z < 0.7$.  This decline is found in
both the Extended Groth Strip (EGS) and the Cosmic Evolution
Survey (COSMOS).  

This evolution is shown in detail in Figure~7 of Conselice et al.
(2009b), in which paper the values we use are also tabulated.  In this paper, we use 
the individually measured merger
fractions in both the COSMOS and EGS surveys, as well as the combination 
of the two surveys, to measure the merger fraction.  We do not discuss
the details of the calculation of the merger fraction, 
although this is described in great detail in  
Conselice (2003), Conselice et al. (2008a) and Conselice et al. (2009a,b).
An important aspect concerning our use of merger fractions
within this paper is that they are measured within the
same rest-frame wavelength, and in the same way, at  different redshifts.

\section{Analysis}

Our analysis first consists of examining how eq. (5) gives an
upper limit on the merger time-scale. We later we utilise eq. (7) to measure
the merger time-scale for our sample while accounting for
any additional mergers that have occurred between $z_1$
and $z_2$.  We carry out our analyses by applying these two
equations to various redshift bins of sizes $\delta z$ = 0.1.  Our 
resulting time-scales
are the average values we find, while the error is the average
error on these measurements, given uncertainties in the measured
merger fractions. We also describe the scatter in these values
at different redshifts.

When we apply equation (5) to our data the maximum time-scale is 
$\tau_{\rm m} = 1.1\pm0.3$ Gyr.  That
is, the maximum time for a CAS asymmetry to last is
roughly $\sim$ 1 Gyr.  This is similar to the maximum time-scale found for
CAS mergers in N-body models by Lotz et al. (2008). This
reveals that any claims that the merger time-scale for CAS
assumed to date (e.g., Conselice 2006a; Lotz et al. 2008) are
underestimated by more than a factor of 2-3 cannot be correct. 

We now utilize the fact that even when the merger fraction is
declining there are still ongoing mergers between different redshifts,
which will raise the merger fraction, and thus result
in an overestimated merger time-scale using eq. (5). We
use equation (7) to calculate the likely true merger time-scale,
through an examination of the merger rate per galaxy, or the
time between mergers, given by $\Gamma$.

Using equation (7) the computed
merger time-scale ranges between 0.3$\pm0.3$ Gyr and
0.9$\pm$0.3 Gyr, with an average value of
$\tau_{\rm m} = 0.6\pm0.3$. In general, we find that the computed
merger time-scale is lower at lower redshifts between
$z = 0.7$ and $z = 0.2$ within the combined EGS and
COSMOS samples.  The values we calculate are
roughly similar to what N-body models of the
merger process find is the time-scale for
CAS sensitivity (e.g., Conselice 2006a; Lotz et
al. 2008).  Our measurement
of merger times is thus likely correct, given the
consistency of independent methods.

Our merger time-scale measurement is smaller by roughly a factor of 
two compared to the results
of Kitzbichler \& White (2008), who calculate merger time-scales
for pairs based on semi-analytical simulations.  Our merger time-scale
is also shorter than other predictions based on semi-analytical
models of merging dark matter halos (e.g., Boylan-Kolchin, Ma \& 
Quataert 2008), suggesting that the implementations of merging
time-scales in these models are off by at least a factor of two
(see also e.g., Conroy et al. 2007; Bertone \& Conselice 2009).

One important caveat about this measurement of the 
merger time-scale is that we cannot assume that it
applies at earlier times, particularly in the early universe
at $z > 2$ when galaxies were more gas rich than they are
today.  In principle, the method used in this paper can
be applied at higher redshifts, but will require more
accurate measures of the merger fraction than that provided
by the available data (e.g., Conselice et al. 2008).  Future
large surveys with WFC3 on Hubble will potentially 
allow this measurement to be made at $z > 1$.

\section{Implications}

\subsection{Number of Major Mergers at $z < 1$}

The implications of this result can be obtained by integrating
the merger rate per galaxy over time to obtain the number of major
mergers a galaxy undergoes since $z \sim 1$.  We assume
in this calculation that the efficiency of detection for
major mergers is 100\% (see Conselice 2006, 2009b). The
number of major mergers is then given by:

\begin{equation}
N_{\rm m} = \int^{t_2}_{t_1} \Gamma^{-1} dt = \int^{z_2}_{z_1} \Gamma^{-1} \frac{t_{H}}{(z+1)} \frac{dz}{E(z)},
\end{equation}

\noindent where $t_{H}$ is the Hubble time, and $E(z) = 
[\Omega_{\rm M}(1+z)^{3} + \Omega_{k}(1+z)^{2} + \Omega_{\lambda}]^{-1/2}$ 
= $H(z)^{-1}$.   Using our average time-scale of 0.6$\pm0.3$ Gyr, we find 
that the total number of mergers occurring since $z = 1.2$ is $N_{\rm m} = 
0.90_{-0.23}^{+0.44}$.  This number is reduced to $N_{\rm m} \sim 0.5$ mergers 
if we use the maximum time-scale of $\tau_{\rm m} = 1.1$ Gyr.   These 
numbers are similar to what we calculate in previous papers in this series
(Conselice et al. 2009b), and even in early work where the merger history at 
$z < 1$ was uncertain (Conselice et al. 2003; Conselice 2006a). However, this 
is the first completely empirical method for measuring the merger history 
at $z < 1.2$, without recourse to models.  

\begin{figure}
 \vbox to 105mm{ 
\includegraphics[angle=0, width=85mm]{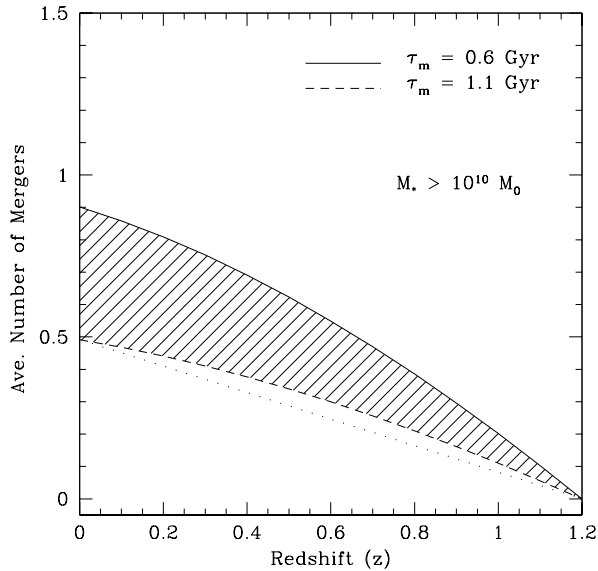}
 \caption{The integrated number of mergers from $z = 1.2$ to
$z = 0$ for galaxies with stellar masses M$_{*} > 10^{10}$ \solm
for two merger time-scale: $\tau_{\rm m} = 0.6$ Gyr and
$\tau_{\rm m} = 1.1$ Gyr.  The shaded area shows the likely
range of possible merger histories, with a range from 0.5
major mergers to a single major merger at $z < 1.2$. }
\vspace{9cm}
} \label{sample-figure}
\end{figure}

This implies that massive galaxies with stellar masses M$_{*} > 10^{10}$ 
\solm, undergo a relatively high number of average major mergers at 
$z < 1.2$.  Most of this merging ($\sim$ 80\%) occurs at $z > 0.5$ (Figure~1).
This shows that the 
major merger process is indeed still occurring, albeit at a reduced rate, for 
these massive systems within the last half of the age of the universe.  

We however still do not know for certain what the mass sensitivity of the 
CAS system is, although N-body models suggest that it is only sensitive to 
mergers which are 1:3 or greater. We use this to calculate how much mass these 
galaxies potentially grow by due to these mergers. If the average
CAS merger is sensitive to a 1:2 major merger, then an average massive galaxy 
with M$_{*} > 10^{10}$ \solm will grow by roughly 50\% since $z = 1.2$.  This 
is similar to the results found by Conselice et al. (2007) who calculated the 
likely amount of merging in massive galaxies at $z < 2$ based on changes in 
the mass function, finding a very similar answer of  0.9$^{+0.7}_{-0.5}$ 
major mergers since $z = 1.4$,  
after removing contributions from star formation (Conselice et al. 2007).

These results also agree with our previous measures based on merger time-scales
calculated from N-body models of the merging process.   We find roughly the same
number of mergers, as described in Conselice et al. (2009b),
when using time-scales calculated through N-body models 
from Lotz et al. (2009).

\subsection{Orbital and Physical Conditions of Major Mergers at $z < 1.2$}

While the results we have discussed and presented thus far concern the
derived merger history, we can also use our empirically
measured time-scales
to determine the likely average physical conditions of the mergers
themselves. This is done by comparing the time-scale we calculate
to the merger-time scales measured through N-body models of equal mass galaxy
mergers from Lotz et al. (2008).  Lotz et al.
have calculated the time-scale for merging as a function of
orbital parameters, and in terms of initial conditions, such
as bulge/disk ratio, and gas mass fraction.  A large fraction
of the differences in Lotz et al. (2008) time-scales is due to the 
orbital parameterisation as one of:
prograde-prograde (SbcPP), prograde-retrograde (SbcPR), retrograde-retrograde
(RR), or prograde-polar (SbcPol). We determine the average merging 
conditions for galaxies within our sample up to $z < 1$, assuming that 
orbital time-scales of the Lotz et al. (2008) major merger models are 
representative of lower mass ratio mergers.

The time-scale for the CAS method within Lotz et al. (2008) ranges
from 0.2 Gyr to 1.5 Gyr. The shorter time-scales are for galaxy mergers which
are not dominated by gas, and have sub-parabolic orbits, which tend to
produce faster merger time-scales.   The longest time-scales occur
for galaxies in the SbcRR model, with a merger
time-scale of nearly $\tau = 1.5$ Gyr (Lotz et al. 2008).  Based
on our time-scale measurement, we can rule out that the average merger
at $z < 1$ is a retrograde-retrograde merger.  Because of their short
time-scales, we can also rule out a rapid merger history in sub-parabolic
orbits.   In fact, the only simple orbital orientation that matches our
average value of the merger time scale of $\tau \sim 0.6$ Gyr is the
prograde-prograde orientation, which has a CAS merger time-scale
in Lotz et al. of $\tau_{\rm m} = 0.74\pm0.17$.  

The best matching orbital configuration, in comparison to our findings,
is the SbcPP model, the SbcPPr- model and the SbcR model.  The SbcPPr-
model is a prograde-prograde merger with a small peri-centric distance,
while the SbcR model is a merger with a highly radial orbit with a
prograde-retrograde orientation.   This reveals that a
typical merging galaxy at $z < 1$ is likely a prograde-prograde
merger. However it remains possible, or even likely, that other merger 
types are occurring, but these cannot dominate the merger process.

\subsection{Could Asymmetries be Produced Through Star Formation?}

In this final subsection we address the question of whether 
the asymmetry criteria for locating mergers could be significant affected 
by star formation events. While it has been shown through using 
the clumpiness index (Conselice 2003),
and a comparison between asymmetries and distorted kinematics (Conselice
et al. 2000b), as well as visual estimates of mergers (Conselice et al.
2005), that ultra-high asymmetries correlate with merging galaxies, we
provide further evidence here based on the asymmetry time-scale.

We argue this based on the fact that the merger time-scale is roughly
$\tau_{\rm m} \sim 0.6$ Gyr, and is no higher than $\sim 1.1$ Gyr at
$z < 1.2$.  If the asymmetric regions in these galaxies were due to star 
forming complexes,
they would last no longer than a few tens of Myr, as the ages of
star formation regions are typically no older than 10-30 Myr (e.g., Palla
\& Galli 1997).  Thus, within roughly half a Gyr, these star forming
regions would no longer be distinct from the rest of the galaxy, and
as such would not stand out when measuring asymmetries. 
We conclude that it is unlikely for star formation to be the
cause of the very high asymmetries we attribute to merging galaxies. 

It is possible that star formation re-occurs throughout our time-scale,
but the drop in the star formation rate is faster than the
derived merger fraction, suggesting the two are not coupled.  For
example, Baldry et al. (2005) find that the star formation rate
declines from 0.15 \solm\,yr$^{-1}$\,Mpc$^{-3}$
at roughly $z = 1$ to $\sim 0.015$ \solm\,yr$^{-1}$\,Mpc$^{-3}$ at
$z \sim 0$. While Conselice et al. (2009b) find that the merger fraction
declines from $f_{\rm m} = 0.13$ at $z = 1.2$ to  $f_{\rm m} = 0.04$
at $z = 0.2$.  While the star formation rate declines by at least
a factor of ten, the merger fraction drops by a factor of three.

\section{Summary}

We have made the first empirical measurement of the time-scale for
mergers within the CAS system, based on the detailed merger
fraction evolution described in Conselice et al. (2009b). These merger
fractions are taken from the Extended Groth Strip and COSMOS surveys,
and constitute $> 20,000$ galaxies with stellar masses M$_{*} > 10^{10}$ \solm.

Our major result is that the time-scale for CAS mergers at $z < 1$ is
between 1.1 Gyr and 0.3 Gyr.  Our best estimated time-scale is 
$\tau_{\rm m} = 0.6\pm0.3$ Gyr, which gives the total number
of mergers occurring at $z < 1.2$ as N$_{\rm m} = 0.90_{-0.23}^{+0.44}$,
similar to previous work based on N-body simulation time-scales, and
from changes in the mass density of galaxies at $z < 1$ (Conselice 
et al. 2007).  We calculate that, on average, a galaxy
with stellar mass M$_{*} > 10^{10}$ \solm will increase its stellar mass
by 50\% due to these mergers.  This time-scale also rules out the
possibility that star formation is the cause of asymmetries seen
in galaxies, as our observed time-scales are over an order of magnitude 
too long to be produced by single star formation events.

The fact that there is a good agreement between empirically derived
merger time-scales and those based on galaxy merger simulations 
suggests that we are beginning to understand the role of mergers
within galaxy evolution. While in the local universe roughly 70\% of
galaxies with masses M$_{*} > 10^{10}$ \solm are disks, the majority
of these contain large bulges, and very few are pure disks
(e.g., Conselice 2006b).  Likely, some of these massive galaxies
are undergoing more evolution than others, and it is possible that
some of the more clustered systems, such as ellipticals are more likely
to undergo more than one merger at $z < 1.2$, which would also help
explain the increase in sizes for these galaxies (e.g., Trujillo
et al. 2007; Buitrago et al. 2008).

These results show that mergers are an important part of
the galaxy formation process at $z < 1.2$, when most galaxies appear
to have morphologies similar to today (e.g., Conselice et al. 2005).
Applying this methodology to higher redshifts will prove more
challenging, due to the active ongoing evolution of these systems
at early times, and the likelihood that some fraction will undergo 
more than a single merger.  This can be probed
in the future when large area surveys for galaxy mergers
 at $z > 1.5$ are carried out.

I thank the referee, Patrik Jonsson, for his comments which 
significantly improved this paper.   I also thank Asa Bluck and Russel White for 
useful conversations on these topics, and support from the STFC.

\appendix

\label{lastpage}


\vspace{-0.5cm}

\begin{thebibliography}{99}
\bibitem[\protect\citeauthoryear{}{}]{b1} Baldry, et al. 2005, MNRAS, 358, 441
\bibitem[\protect\citeauthoryear{}{}]{b1} Bertone, S., Conselice, C.J. 2009, MNRAS, arXiv:0904.2365
\bibitem[\protect\citeauthoryear{}{}]{b1} Bluck, A.F.L., Conselice, C.J., Bouwens, R.J., Daddi, E., Dickinson, M., Papovich, C., Yan, H. 2009, MNRAS, 394, 51L 
\bibitem[\protect\citeauthoryear{}{}]{b1} Boylan-Kolchin, M., Ma, C.-P., Quataert, E. 2008, MNRAS, 383, 93
\bibitem[\protect\citeauthoryear{}{}]{b1} Buitrago, F., Trujillo, I., Conselice, C.J., Bouwens, R.J., Dickinson, M., Yan, H. 2008, ApJ, 687, 61L
\bibitem[\protect\citeauthoryear{}{}]{b1} Conselice, C.J., Bershady, M.A., Jangren, A. 2000a, ApJ, 529, 886
\bibitem[\protect\citeauthoryear{}{}]{b1} Conselice, C.J., Bershady, M.A., Gallagher, J.S. 2000b, A\&A, 354, 21L
\bibitem[\protect\citeauthoryear{}{}]{b1} Conselice, C.J. 2003, ApJS, 147, 1
\bibitem[\protect\citeauthoryear{}{}]{b1} Conselice, C.J., Bershady, M.A., Dickinson, M., Papovich, C. 2003, AJ, 126, 1183
\bibitem[\protect\citeauthoryear{}{}]{b1} Conselice, C.J., Blackburne, J., Papovich, C. 2005, ApJ, 620, 564
\bibitem[\protect\citeauthoryear{}{}]{b1} Conselice, C.J. 2006a, ApJ, 638, 686
\bibitem[\protect\citeauthoryear{}{}]{b1} Conselice, C.J. 2006b, MNRAS, 373, 1389
\bibitem[\protect\citeauthoryear{}{}]{b1} Conselice, C.J., et al. 2007, MNRAS, 381, 962
\bibitem[\protect\citeauthoryear{}{}]{b1} Conselice, C.J., Rajgor, S., Myers, R. 2008, MNRAS, 386, 909 (paper I)
\bibitem[\protect\citeauthoryear{}{}]{b1} Conselice, C.J., Arnold, J. 2009a, MNRAS, arXiv:0904.4250 (Paper II)
\bibitem[\protect\citeauthoryear{}{}]{b1} Conselice, C.J., Yang, C., Bluck, A.F.L. 2009b, arXiv:0812.3237 (Paper III)
\bibitem[\protect\citeauthoryear{}{}]{b1} Conroy, C., Ho, S., White, M. 2007, MNRAS, 379, 1491
\bibitem[\protect\citeauthoryear{}{}]{b1} Kitzbichler, M.G., White, S.D.M. 2008, MNRAS, 391, 1489
\bibitem[\protect\citeauthoryear{}{}]{b1} Lotz, J.M., Jonsson, P., Cox, T.J., Primack, J.R. 2008, MNRAS, 391, 1137
\bibitem[\protect\citeauthoryear{}{}]{b1} Palla, F., Galli, D. 1997, ApJ, 476, 35L
\bibitem[\protect\citeauthoryear{}{}]{b1} Trujillo, I., Conselice, C.J., Bundy, K., Cooper, M.C., Eisenhardt, P., Ellis, R.S. 2007, MNRAS, 382, 109
\end{thebibliography}
\end{document}